\documentclass[11pt, onecolumn,epsf]{IEEEtran}
\usepackage{graphicx, epstopdf,cite,subfigure,cite}
\usepackage{amsthm,amsmath,amssymb}
\usepackage{multirow}
\usepackage{url}
\usepackage{graphics}
\usepackage{array}
\usepackage{color,soul}
\usepackage{url, cite}

\begin{document}

\input epsf
\title {Beam with Adaptive Divergence Angle in Free-Space Optical Communications for High-Speed Trains}

\author{{Yagiz~Kaymak,~\IEEEmembership{Student Member,~IEEE}, Sina Fathi-Kazerooni, Roberto~Rojas-Cessa,~\IEEEmembership{Senior Member,~IEEE}, JiangHua~Feng, Nirwan Ansari,~\IEEEmembership{Fellow,~IEEE}, MengChu Zhou,~\IEEEmembership{Fellow,~IEEE}, and Tairan~Zhang \thanks{Y. Kaymak, S. F. Kazerooni, R. Rojas-Cessa, N. Ansari, and M. Zhou are with the Department of Electrical and Computer Engineering, New Jersey Institute of Technology, Newark, NJ 07102. Email: \{{yk79, sf267, rojas, ansari, zhou}\}@njit.edu.}
\thanks{M. Zhou is also with the Institute of Systems Engineering, Macau University of Science and
Technology, Macau 999078, China.}
}\thanks{J. Feng and T. Zhang are with CRRC Zhuzhou Institute Co., Ltd, Shidai Road, Zhuzhou, Hunan Province, China. Email: {\{fengjh, zhangtrg\}@csrzic.com.}}}

\date{August-25-2018}
\maketitle \thispagestyle{empty}

\begin{abstract}
In this paper, we propose an adaptive beam that adapts its divergence angle according to the receiver aperture diameter and the communication distance to improve the received power and ease the alignment between the communicating optical transceivers in a free-space optical communications (FSOC) system for high-speed trains (HSTs). We compare the received power, signal-to-noise ratio, bit error rate, and the maximum communication distance of the proposed adaptive beam with a beam that uses a fixed divergence angle of 1 mrad. The proposed adaptive beam yields a higher received power with an increase of 33 dB in average over the fixed-divergence beam under varying visibility conditions and distance. Moreover, the proposed adaptive divergence angle extends the communication distance of a FSOC system for HSTs to about three times under different visibility conditions as compared to a fixed divergence beam. We also propose a new ground transceiver placement that places the ground transceivers of a FSOC system for HSTs on gantries placed above the train passage instead of placing them next to track. The proposed transceiver placement provides a received-power increase of 3.8 dB in average over the conventional placement of ground-station transceivers next to the track.
\end{abstract}

\begin{IEEEkeywords}
free-space optical communications, optical wireless communications, high-speed trains, beam divergence, adaptive divergence angle.
\end{IEEEkeywords}

\section{Introduction}
\label{sec:intro}
\IEEEPARstart{H}{igh-speed} trains (HSTs) are an essential means of public transportation for millions of people all around the world. As an example, 127 million passengers used the Beijing-Shanghai high-speed rail in 2016 \cite{chen2016impact}. According to Worldwide Railway Organization, the high-speed railroad traffic in the world has increased from 248.2 billion passenger-kilometer (i.e., the distance in kilometers traveled by a passenger) in 2010 to 715.7 billion passenger-kilometer in 2016, which corresponds to a 188\% hike in HST traffic \cite{hst}.

The demand for broadband Internet access on board by high-speed train passengers has been also growing with the proliferation of smart phones and other mobile devices. These passenger equipment is usually used for running online applications, streaming on demand or live video, videoconferencing, online messaging, and web browsing while on-board \cite{masson2015broadband}. Considering the data rates required by those applications and that the number of passengers on a single HST ranges from 500 to 1300 \cite{highSpeedPassengers}, an aggregated demand is on the order of Gbps. For instance, a YouTube video with a resolution of 1280x720 pixels (i.e., 720p) requires a recommended bitrate of 2,500 Kbps \cite{youtube_bitrates}. Therefore, a video demand generated by 500 passengers, each receiving a 720p video stream, at the same time would require an aggregated rate of 1.25Gbps. 

There are several technologies, such as Wi-Fi \cite{yamada2010communication},  long-term evolution (LTE) \cite{parichehreh2016seamless}, WiMAX \cite{aguado2008wimax}, and Radio-over-fiber (RoF) \cite{lannoo2007radio} being considered for HST communications. However, providing broadband Internet access (i.e., in the order of Gbps) to the passengers on a HST by using existing radio frequency (RF) communications systems is challenging because of the high traveling speed of the train and the limited bandwidth available to RF technologies \cite{zhou2011broadband}. RF communications technologies may fall short of providing high data rates to HSTs because of the impact of the Doppler effect, frequent handovers, and operational frequencies and bandwidths \cite{ericsson, kaltenberger2015broadband}. The upcoming 5G communications technology, using millimeter wave, may be also employed for HST communications \cite{wang2014cellular}. It is expected that 5G will provide a peak data rate of 10 Gbps in low mobility scenarios, such as for local wireless access, and 1 Gbps in high mobility scenarios in the near feature \cite{wang2014cellular}. However, 5G is not yet deployed and requires spectrum licensing. Free-space optical communications (FSOC), which is also known as optical wireless communications (OWC), is a line-of-sight (LOS) technology that propagates modulated light to transmit data between stations in stationary or mobile conditions \cite{alkholidi2014free}. Recently, FSOC technology has attracted considerable attention because it has the potential to provide transmissions at very-high data rates between two terminals separated over a distance that varies from a few meters to thousands of kilometers. FSO finds its applicability to stationary and mobile scenarios including building-to-building communications, HSTs, unmanned aerial vehicles (UAVs), satellites, indoor and outdoor local- and wide-area networks, and deep space communications \cite{kaymak2018survey}. FSOC possess multiple advantages, such as high bandwidth, license-free band use, long operational range, spatial re-usability, security, and immunity to electromagnetic interference as compared to existing RF communications systems \cite{khalighi2014survey}. Frequencies used by FSOC are much higher than those used by RF communications. Moreover, these high data rates can be achieved while using antennas that occupy a small footprint \cite{edwards2014nasa}. Moreover, the coherence of laser lights employed in FSOC systems as light sources may focus the beam within a small area and enable the transmission of high data rates at long distances \cite{kim2001comparison, KaushalK15}. 

A laser beam emitted from a laser diode diverges as it exits the emitting aperture in the active layer of the laser diode because of the resulting diffraction of the light waves that occurs as the light is bent by the corners of the emitting aperture \cite{svelto1998principles, KaushalK15}. Diffraction causes light waves to spread transversely as they propagate, and it is, therefore, impossible to generate a perfectly collimated beam \cite{gaussian_beam_optics}. The divergence of a beam, which is measured as the divergence angle of the transmitted beam, increases the beam width as the distance from its source increases. Therefore, a beam with a fixed divergence angle may illuminate an area larger than the receiver aperture area of a distant receiver. In this way, the receiver aperture may only collect a fraction of the incident beam. The remaining uncollected light results in a loss called divergence or geometric loss \cite{alkholidi2014free}. The geometric loss decreases the received power if the other communication parameters, such as the communication distance, transmission power, and the diameter of the transmitting and receiving apertures remain constant. The received power in a FSOC system may be increased by using a beam width smaller than or equal to the diameter of the receiver aperture for a given communication distance. By selecting a beam divergence angle that generates a beam width smaller than the diameter of a receiver aperture, the required alignment accuracy between the communicating terminals increases. In contrast, the use of a fixed divergence beam may not produce an optimum received power with ease of alignment because of the varying communication distance between a base station and the train as the train travels. 

Therefore, we propose an adaptive divergence approach that adapts the beam divergence angle of the transmitted beam to achieve a footprint of the diameter of the receiver aperture for a given communication distance between transmitter and receiver. We also propose to place ground transceivers of a FSOC system for HSTs top of the track (right above the passage of the train) to achieve an efficient alignment. This new placement may improve the received power by decreasing the lateral distance between the train and the ground transceivers, and makes the ground transceivers parallel to the track. The proposed transceiver placement provides a received-power increase of 3.8 dB in average over the conventional transceiver placement next to track \cite{paudel2013modelling, kaymak2017, fathi2017optimal}. We elaborate the system model of the proposed ground transceiver placement in Section \ref{sec:systemModel}.

The remainder of the paper is organized as follows. Section \ref{sec:systemModel} presents our system model. Section \ref{sec:results} presents numerical results of our adaptive divergence angle model. Section \ref{sec:relatedwork} summarizes the related work.  Section \ref{sec:conclusions} concludes the paper.

\section{System Model}
\label{sec:systemModel}
\begin{figure}[t!]
	\centering
	\includegraphics[width=3.5in]{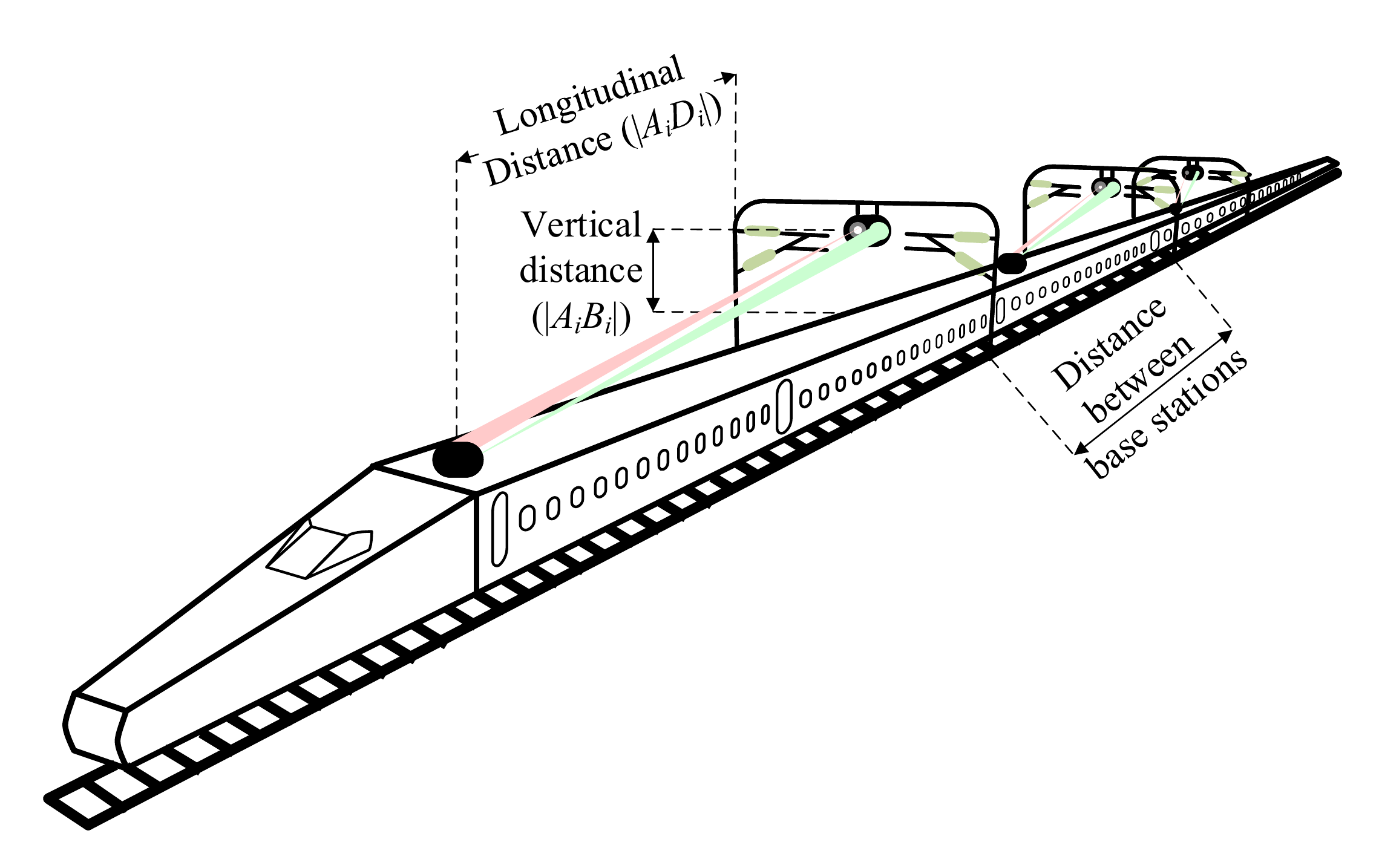}
	\caption{3D view of the proposed FSOC system for HSTs.}
	\label{fig:trainmodel}
\end{figure}

\begin{figure}[t!]
	\centering
	\includegraphics[width=3.5in]{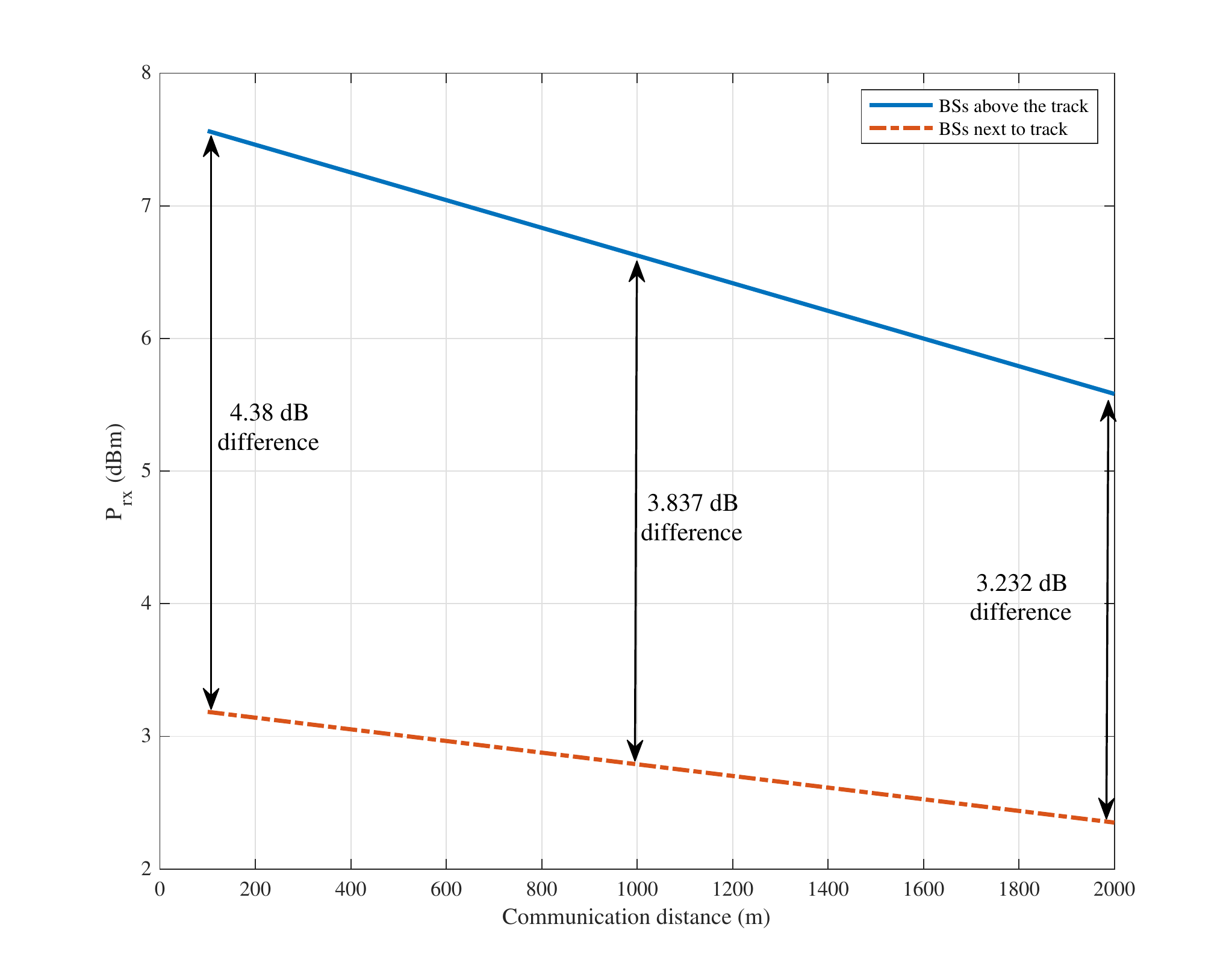}
	\caption{Comparison of received power for a system that places base stations on the track's side and another above the track, both using adaptive divergence angle. The visibility is 5 km and transmission power is 10 dBm.}
	\label{fig:model_comparison}
\end{figure}

Several geometric models of FSOC systems for HSTs \cite{paudel2013modelling, kaymak2017, fathi2017optimal} that place the base stations next to track have been proposed. Among these studies, the results in \cite{fathi2017optimal}  showed that decreasing the lateral distance, which corresponds to the vertical distance in our system model, between a base station and the train when the base stations are placed next to the track improves the received power and increases the coverage area of the base station. The results in \cite{fathi2017optimal} provide a motivation to propose a FSOC system for HSTs, where the base stations are placed on top of the track (and above the train passage), as shown in Figure \ref{fig:trainmodel}. In this case, gantries may be used as the supporting structure for ground transceivers. The FSOC system having the base stations attached to gantries instead of placing them next to the track decreases the lateral distance between the base stations and the train, thus, improving the received power, signal-to-noise ratio (SNR), and BER. Figure \ref{fig:model_comparison} compares the received power of a FSOC model that places the base stations next to the track \cite{kaymak2017} and our proposed ground transceiver placement where the base stations are attached to gantries. Both models in Figure \ref{fig:model_comparison} use an adaptive beam that adjusts its divergence angle according to the diameter of the receiver aperture and the communication distance. According to the results in this figure, the proposed placement achieves a higher received power with an increease of 3.8 dB on average. 

Each base station in the proposed ground transceiver placement is attached to a gantry that may be used in the power network along the track. Therefore, this placement may not only increase the data rates or inter-station distance but also it may decrease deployment costs. Moreover, at least two transceivers are attached to a HST in the proposed model, which allows establishing multiple optical connections between the train's transceivers and multiple consecutive base stations. Establishing multiple simultaneous optical connections between the train and the consecutive base stations may improve the reliability and increase the aggregated data rate of the proposed FSOC system for a train passing through an area covered by multiple base stations. For instance, the three transceivers on the train in Figure \ref{fig:trainmodel} are connected to three consecutive base stations, and this multi-link connection may improve the reliability of the proposed FSOC system.

\subsection{Geometric Model}
\label{subsec:geometricModel}
\begin{figure}[t!]
	\centering
	\includegraphics[width=3.5in]{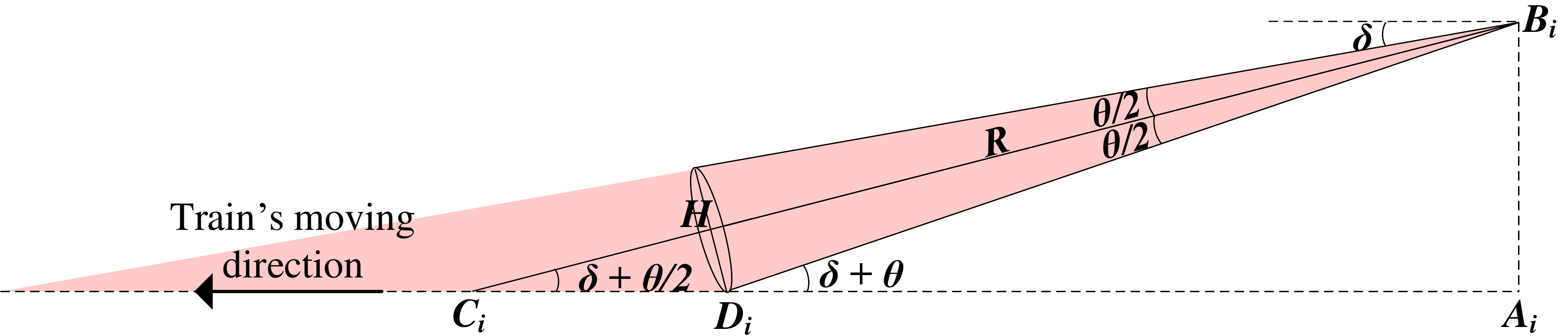}
	\caption{Geometric representation (lateral view) of the transmitted beam from a base station located at $B_i$.}
	\label{fig:geometricmodel}
\end{figure}

The geometric model of the proposed FSOC system presented in this section enables the calculation of the received power in Section \ref{subsec:linkbudget}. Figure \ref{fig:geometricmodel} shows the lateral view of a beam transmitted from a base station located at $B_i$. The transmitted beam is received by a transceiver on the train located at $H$. 

The longitudinal distance between the train's transceiver at $H$ and a base station at $B_i$ is denoted as $|A_i D_i|$. The vertical distance between the train and the base station is $|A_i B_i|$. Therefore, $|B_i D_i|$ can be calculated as 

\begin{equation}
\centering
|B_i D_i| = \sqrt{{|A_i B_i|}^2 + {{|A_i D_i|}^2}}
\label{eq:BiDi}
\end{equation}

The radius of the receiver aperture, $r_{rx}$, is equal to $\sqrt{S_{rx}/\pi}$ for a given receiver aperture area, $S_{rx}$. Therefore, the divergence angle of the transmitted beam is calculated as
 
 \begin{equation}
 \centering
 \theta_{1/2} = \arcsin{\bigg(\frac{r_{rx}}{|B_i D_i|}\bigg)}
 \label{eq:BiDi}
 \end{equation}
 
 The communication distance between the transmitter and receiver, $R$, (in Figure \ref{fig:geometricmodel}) is calculated as
 
  \begin{equation}
 \centering
 R = \cos{(\theta_{1/2})}|B_i D_i|
 \label{eq:R}
 \end{equation}

\subsection{Calculation of Received Power and Impact of Fog}
\label{subsec:linkbudget}
The received power of the optical radiation along the axis of propagation is calculated according to Friis formula \cite{majumdar2010free}:

\begin{equation}
\centering
P_{rx} = P_{tx} G_{tx} G_{rx} {\Big(\frac{\lambda}{4 \pi R}\Big)}^2 L_{geo} L_{tx} L_{rx} \eta_{tx} \eta_{rx}
\label{eq:Prx}
\end{equation}
where $P_{rx}$, $P_{tx}$, $G_{tx}$, $G_{rx}$, $\lambda$, $R$, $L_{geo}$, $L_{tx}$, $L_{rx}$, $ \eta_{tx}$, and $\eta_{rx}$ are the received power, transmission power, transmitter antenna gain, receiver antenna gain, wavelength of the beam, communication distance between the transmitter and receiver, geometric loss, transmitter pointing loss, receiver pointing loss, transmitter optical efficiency, and the receiver optical efficiency, respectively.

In general, laser-beam propagation can be approximated by assuming that the lasers emit beams with a Gaussian profile, where the laser is said to be operating on the fundamental transverse mode, or TEM00 mode of the laser's optical resonator \cite{andrews2005laser}. Therefore, we follow this approximation and assume that the laser beam used in this work has a Gaussian profile. The approximation of the transmitter antenna gain for a Gaussian beam is given by \cite{lambert1995laser}:

\begin{equation}
\centering
G_{tx} = \frac{32}{\theta^2}
\label{eq:Gt}
\end{equation}
where $\theta$ is the divergence angle of the transmitted beam in radians. The receiver antenna gain is given by \cite{majumdar2010free, henniger2010introduction}:

\begin{equation}
\centering
G_{rx} = \Big(\frac{\pi D_{rx}}{\lambda}\Big)^2
\label{eq:Gr}
\end{equation}
where $D_{rx}$ is the telescope aperture diameter, in meters. The laser beam disperses conically upon exiting the transmitter lens. This dispersion increases as the distance from the laser source increases according to the geometrical loss, $L_{geo}$, which is given by \cite{bloom2003understanding}:

\begin{equation}
\centering
L_{geo} =  \Big(\frac{D_{rx}}{D_{tx} + \theta R}\Big)^2
\label{eq:Lgeo}
\end{equation}
where $D_{tx}$ is the diameter of the transmitter, in meters.

$L_{tx}$ and $L_{rx}$ in (\ref{eq:Prx}) are the transmitter and receiver pointing loss \cite{liu2009free}, respectively, which are given by 

\begin{equation}
\begin{aligned}
L_{tx} = e^{-G_{tx} \gamma^2},
\\
L_{rx} = e^{-G_{rx} \zeta^2}
\label{eq:Lt_Lrx}
\end{aligned}
\end{equation}
where $\gamma$ and $\zeta$ denote the radial pointing errors of the transmitter and receiver in radians, respectively.

Fog and rain attenuate the propagating beam as the water molecules/droplets of fog and rain absorb and scatter the optical signal \cite{alkholidi2014free}. Fog is the most dominant atmospheric attenuating factor for FSOC systems among all weather conditions as the radius of water molecules of fog is in the range of the wavelength of the communicating beam \cite{alkholidi2014free}. Therefore, we take the impact of fog into consideration in calculating the received power. Some empirical fog models, such as Kruse \cite{kruse1962elements}, Kim \cite{kim2001comparison}, and Ijaz \cite{ijaz2013modeling} have been proposed to express the fog-induced power loss in dB/km \cite{ijaz2013modeling}. These fog models represent the received power as a function of the transmission wavelength, meteorological visibility, and the coefficient related to the particle size distribution in the atmosphere. Kruse model is considered to be not accurate enough for visibilities less than 0.5 km since it is originally proposed for haze particles \cite{ijaz2013modeling}. Moreover, Kim model does not take the relationship between visibility and wavelength into account for visibilities less than 0.5 km. Therefore, we adopt Ijaz fog model for visibilities in the range of  [0.015, 1) km. Kim model is adopted for visibilities greater than 1 km. Ijaz and Kim fog models share a common loss function that is given by \cite{ijaz2013modeling, kim2001comparison}:

\begin{equation}
\centering
L_a = \frac{17}{V}{\Big(\frac{\lambda}{0.55 \mu m}\Big)}^{-q(\lambda)}\label{eq:fog}
\end{equation}
where $L_a$ is in dB/km, $V$ is the meteorological visibility in km, $\lambda$ is the transmission wavelength of the laser in $\mu$m, and $q$ is the size distribution of the scattering particles. According to the meteorological visibility, $q$ values are given by \cite{ijaz2013modeling, kim2001comparison}:

\begin{equation}
\centering
q = \begin{cases} 
1.6, & V \geq 50\ km \\
1.3, & 6 \leq V< 50\ km \\
0.16V + 0.34, & 1 \leq V< 6\ km \\
0.1428\lambda - 0.0947, & 0.015 < V < 1\ km \\
0, & V\leq 0.015\ km
\end{cases}
\label{eq:fog_q}
\end{equation}

The received power after the impact of fog is calculated by subtracting the fog-induced power loss,  (\ref{eq:fog}), from the received power calculated in (\ref{eq:Prx}), which is given by

\begin{equation}
\centering
P_{rx_{fog, dBm}} =  10\log_{10}P_{rx} - L_a 
\label{eq:Prx_fog}
\end{equation}

\subsection{Detection of Optical Radiation}
\label{subsec:detectionofOpticalRadiation}
\begin{figure}[t!]
	\centering
	\includegraphics[width=3.5in]{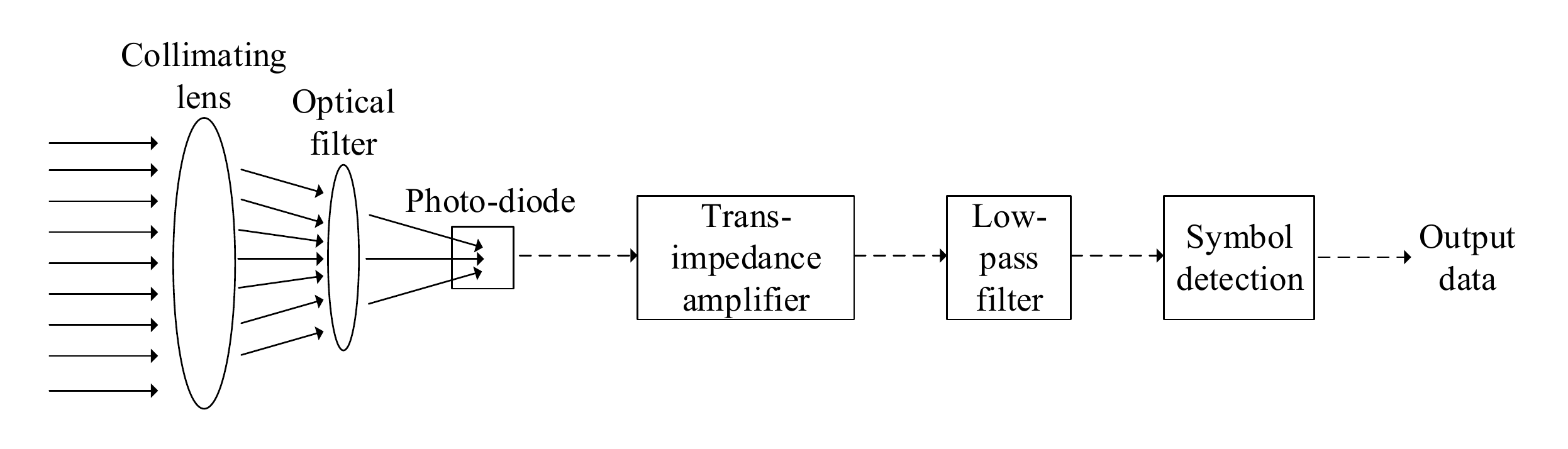}
	\caption{Block diagram of an FSO receiver.}
	\label{fig:blockdiagramofopticalreceiver}
\end{figure}

The optical radiation transmitted from a base station is received by a direct detection receiver on the train. Each direct-detection FSO receiver in the proposed communications system consists of a collimating lens that collects and focuses the incident light, an optical filter to filter out the undesirable background radiations such as direct, reflected, or scattered sunlight, a photodiode that converts the optical signal to electrical signal, a trans-impedance amplifier to amplify the electrical signal, a low-pass filter to limit the thermal and background noise, and a symbol detector to recover the received data. Figure \ref{fig:blockdiagramofopticalreceiver} shows the general block diagram of such a direct-detection optical receiver.

The receiver equipment shown in Figure \ref{fig:blockdiagramofopticalreceiver} may induce some noise that degrades SNR at the receiver. The total noise for a direct detection receiver that employs an avalanche photodiode (APD) is a combination of the photo-current shot, thermal (i.e., Johnson noise), dark current, and background illumination noises \cite{kahn1997wireless, trisno2006design}. The thermal noise, also called Johnson or Nyquist noise, is the electronic noise induced by the thermal agitation of the electrons passing through an electrical conductor \cite{trisno2006design}. The dark current is the current that flows through the bias circuit of a photodiode even without the incident light \cite{trisno2006design}. The dark current noise arises from electrons and/or holes that are thermally generated in the p-n junction of a photodiode. The background noise is the result of the undesirable background radiation collected by the photo detector, which may arise from the intense and visible background light, such as the sun light and artificial lights \cite{kahn1997wireless, manor2003performance}. The shot noise, which is also known as the quantum noise in optical communications, originates from the random occurrence of photon absorption events in a photo detector \cite{manor2003performance}. The number of photons of the incident light fluctuates by following a Poisson distribution at the receiver causes the shot noise. If all noise sources have zero mean and are statistically independent of each other, then the total noise power at the receiver can be calculated by

\begin{equation}
\centering
\sigma_{total}^2 = \sigma_{dark}^2 + \sigma_{background}^2 + \sigma_{shot}^2 + \sigma_{thermal}^2
\label{eq:totalNoise}
\end{equation}

The average received power is typically larger than the signal current, which makes dark current and background noises negligible in practice \cite{trisno2006design}. Therefore, the total variance of the receiver-related noise of a receiver that employs an APD is denoted by

\begin{equation}
\centering
\sigma_{total_{APD}}^2 = \sigma_{shot}^2 + \sigma_{thermal}^2
\label{eq:apdNoise}
\end{equation}
where $\sigma_{shot}^2$ of an APD is given by

\begin{equation}
\centering
\sigma_{shot}^2 = 2 q M i_M F(M) \Delta f
\label{eq:shotNoise}
\end{equation}
where $q$ is the electron charge, $M$ is the multiplication gain of the APD, $i_M$ is the average value of total multiplied output current, $F(M)$ is  the excess noise factor of the photodiode, and $\Delta f$ is the system bandwidth. The multiplication gain of an APD, $M$, is a statistical process and defines the ratio between the multiplied output photocurrent, $i_M$, and the primary un-multiplied photocurrent, $i_S$ \cite{trisno2006design}. The excess noise factor, $F(M)$, of an APD results in an increase in the statistical noise caused by the multiplication process. In other words, $F(M)$ is a multiplier indicating the increase in noise if all the photo-carriers of an APD were multiplied by $M$. $\Delta f$ is the system bandwidth, which is typically selected as the data rate of the FSOC system.

Thermal noise, $\sigma_{thermal}^2$, of an APD is given by
\begin{equation}
\centering
\sigma_{thermal}^2 = \frac{4 k T \Delta f}{R_{load}}
\label{eq:thermalNoise}
\end{equation}
where $k$ is the Boltzmann constant, $T$ is the absolute temperature, and $R_{load}$ is the load resistance of the trans-impedance amplifier.

The SNR, which is the ratio of the average signal power over the average noise power, at the receiver is given by \cite{kahn1997wireless}:
\begin{equation}
\centering
SNR = \frac{({SP_{rx_{fog, watt}}})^2}{\sigma_{total_{APD}}^2}
\label{eq:SNR}
\end{equation}
where $S$ is the photodiode sensitivity, in A/W, and $P_{rx_{fog, watt}}$ and $\sigma_{total_{APD}}^2$ are calculated as in (\ref{eq:Prx_fog}) and (\ref{eq:apdNoise}), respectively. We consider ON-OFF keying (OOK) -non-return-to-zero (NRZ) as the widely adopted modulation scheme \cite{bloom2003understanding, KaushalK15}. The BER of a FSOC system that uses an OOK-NRZ modulation is calculated by

\begin{equation}
\centering
BER = Q(\sqrt{SNR})
\label{eq:BER}
\end{equation}
where $Q$ function is given by \cite{karagiannidis2007improved, gradshteyn2014table}:

\begin{equation}
\centering
Q(x) = \dfrac{1}{\sqrt{2\pi}}\int_{x}^{\infty}e^{-u^{2}/2}du
\label{eq:Qfunc}
\end{equation}

\subsection{Beam Divergence Adjusting Mechanisms}
\label{subsec:BeamDivergenceControl}
In this section, we introduce some beam divergence adjusting mechanisms to obtain a desired beam divergence for the proposed adaptive beam. 

A beam expander is an optical device that accepts a collimated beam as the input and expands the diameter of the beam as the beam leaves the expander \cite{heng2008adaptive}. A beam expander may also reduce the beam width if the expander is used in the reverse way. Note that the collimated beam diameter is proportional to the divergence angle of a beam. Therefore, a beam expander may alter the beam divergence of an input beam by changing its diameter. A simple beam expander, which is also referred to as Keplerian telescope, consists of two lenses with different diameters and focal lengths. The magnification ratio of a Keplerian beam expander is equal to the ratio of the focal lengths of the employed lenses. For instance, the magnification ratio of a beam expander with two lenses having the focal lengths of $f_1$ and $f_2$, respectively, is equal to $\frac{f_2}{f_1}$. Another implementation of beam expansion can be realized by using a Galilean telescope that also uses two lenses, one with positive and the other with negative focal length.

A motorized beam expander, which usually incorporates groups of moving lenses, adjusts the diameter and hence, the divergence angle of the output beam within its magnification range \cite{altechna_motorized_beam_expander, newport_beam_expanders}. Therefore, by using a motorized beam expander the divergence of a beam may be adjusted dynamically. Another method to adjust the beam divergence is to use a 1xN optical switch with one input and N output ports \cite{heng2008adaptive}. The input port of the optical switch is connected to a laser diode or a fiber optic cable that generates the input beam. Each output port of the optical switch is connected to a beam expander with a different magnification ratio. The desired divergence angle of the output beam is obtained by the selection of a specific output port that is connected to the beam expander. The motorized-beam-expander approach yields a continuously-variable divergence angle for the output beam. On the other hand, the approach that uses a 1xN optical switch may allow adjustment of the divergence angle of the output beam discretely because one beam expander with a fixed magnification ratio can be selected at a time by forwarding the inbound optical beam to only one output port of the optical switch. A third method to adjust the divergence angle of a beam is a combination of the first two approaches, where a 1xN optical switch with N motorized beam expanders connected to the switch's output ports is employed. This beam-divergence adjustment method is the most flexible one in terms of the magnification ratio owing to the various magnification ranges of the motorized beam expanders connected to the switch. A beam-divergence adjustment method that uses a motorized beam expander, however, may induce a delay called the expansion change time/delay, which is the time required to move the lenses of a beam expander to create the desired output beam. Some commercial beam expanders have expansion change times of less than 5 seconds \cite{altechna_motorized_beam_expander}. Therefore, we prefer to use a 1xN optical switch with N beam expanders that have fixed magnification ratios to avoid the expansion change delay in this work.

\section{Results and Discussion}
\label{sec:results}
In this section, we provide the numerical analysis performed in MATLAB\textsuperscript{\textregistered}, where the adaptive and fixed divergence angle approaches are compared in terms of the received power, communication distance, SNR, and BER. Table \ref{table:parameters} shows the parameters used in the numerical analysis.

\begin{table}[]
	\centering
	\caption{Evaluation parameters}
	\label{table:parameters}
	\begin{tabular}{|c|c|c|c|}
		\hline
		\textbf{Parameter}       & \textbf{Symbol} & \textbf{Value}                      & \textbf{Unit} \\ \hline
		Wavelength        & $\lambda$                   & 1550               & nm                  \\ \hline
		Transmission power        & $P_{tx}$                   & 10               & mW                  \\ \hline
		Photodiode sensitivity & $S$                  & 0.9                                     & A/W                     \\ \hline
		Electronic charge        & $q$                   & 1.602x10$^{-19}$               & C                  \\ \hline
		Boltzmann constant     &$k$                   & 1.38x10$^{-23}$                & J/K               \\ \hline
		Absolute temperature & $T$                  & 298                                      & K                      \\ \hline
		Multiplication gain of the APD & $M$                  & 10                                      & -                      \\ \hline
	   Excess noise factor & $f(M)$                  & 3.2                                      & -                     \\ \hline
	   System bandwidth & $\Delta f$                  & 10$^9$                                     & Hz                     \\ \hline
	  Resistance of the amplifier & $R_{load}$                  & 50                                     & Ohm                     \\ \hline
	   Surface area of the transmitter & $S_{rx}$                  & 9                                     & cm$^2$                    \\ \hline
	   Surface area of the receiver & $S_{rx}$                  & 95                                     & cm$^2$                    \\ \hline
	   System losses & $L_{sys}$                  & 0.5                                   & -                    \\ \hline
	\end{tabular}
\end{table}

\begin{figure}[t!]
	\centering
	\includegraphics[width=3.5in]{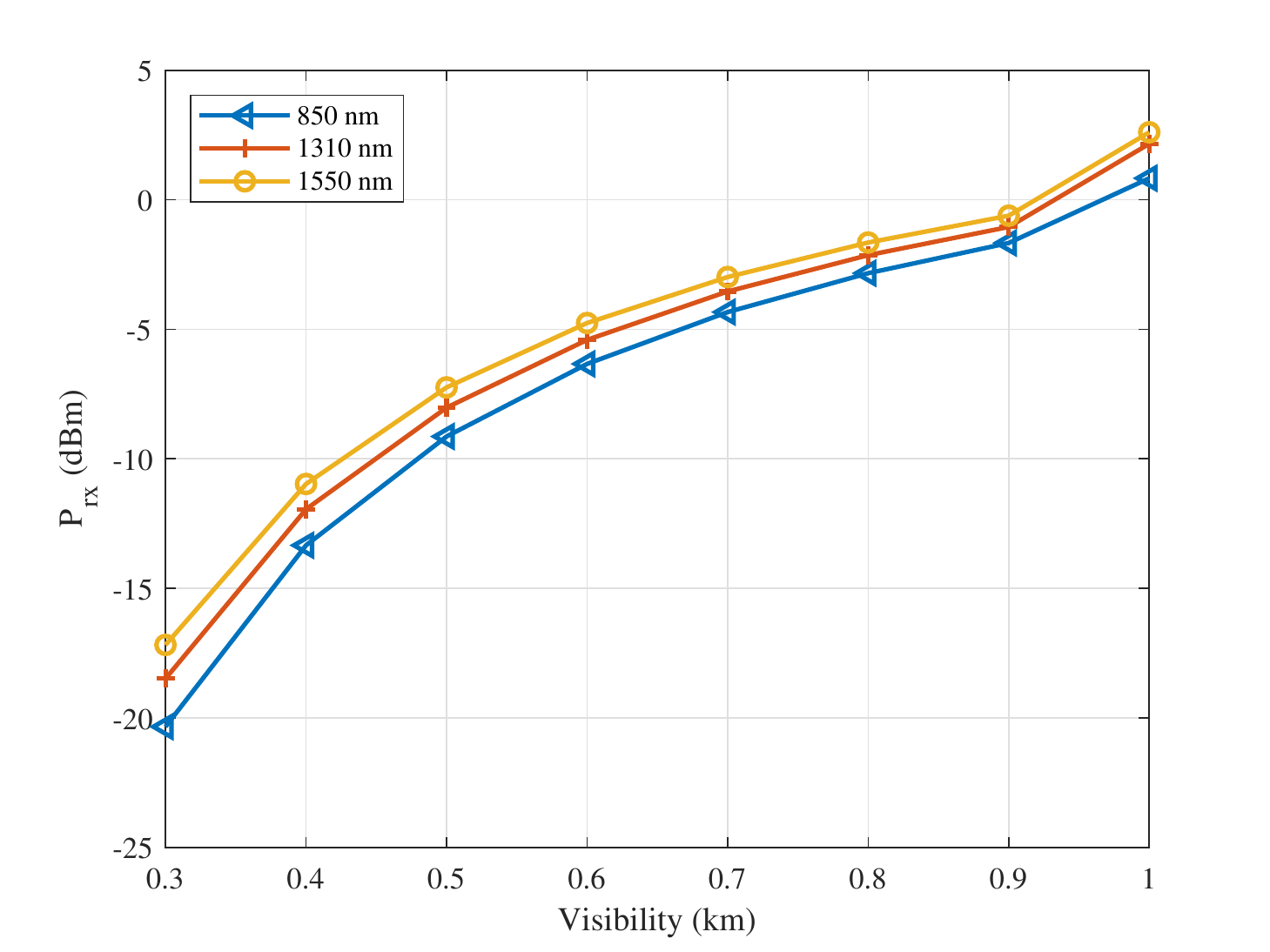}
	\caption{Impact of the wavelength on the received power as visibility varies.}
	\label{fig:wavelength_dependency}
\end{figure}

We use a laser diode transmitting at the wavelength of 1550 nm as the light source. 1550 nm is  a well-studied wavelength for FSOC with the following advantages: 1) It falls in one of the atmospheric windows where the atmospheric attenuation is low. 2) High quality transmitter and detector components that use 1550-nm wavelength are available in the market and they are capable of transmitting high power (i.e., more than 500 mW) and high data rates (i.e., more than 2.5 Gbps). 3) Lasers that use a wavelength of 1550 nm can transmit 50 to 65 times the transmission power of the lasers transmitting at 780 to 850 nm for the same eye safety classification \cite{bloom2003understanding, kim2001comparison}. Figure \ref{fig:wavelength_dependency} compares the impact of transmission wavelength on the received power for the three most-common wavelengths, 850, 1310, and 1550 nm that are used for FSOC as the visibility varies. In Figure \ref{fig:wavelength_dependency} the transmission power and the communication distance are 10 dBm and 500 m, respectively. The 1550-nm light (line with circular marks) yields the highest received power values among the represented wavelengths for the visibilities ranging from 0.3 to 1 km. These results support our wavelength selection for the proposed FSOC system. 

The transmission power of the light source used in the evaluations is selected as 10 mW to make the light source eye safe. Specifically, a laser transmitting at 1550 nm with a transmission power of 10 mW is considered as a Class 1 laser, which is eye safe in an exposure of for up to 100 seconds \cite{ulsm}. The sensitivity, multiplication gain, and the excess noise factor of the selected photodiode given in Table \ref{table:parameters} are typical values for high-speed APDs \cite{nakajima2016high}. The resistance of the load resistor of the trans-impedance circuitry is selected as 50 $\Omega$, which is suitable for high rate FSO links. The transmitter and receiver telescopes with the surface areas of 9 and 95 cm$^2$, respectively, are available in the market \cite{large_plano_convex}. We combine $L_{tx}$, $L_{rx}$, $ \eta_{tx}$, and $\eta_{rx}$ to derive  the system loss, denoted by $L_{sys}$ in Table \ref{table:parameters}. $L_{sys}$ = 0.5 is used in the evaluations \cite{fathi2017optimal}.

\begin{figure}[t!]
	\centering
	\includegraphics[width=3.5in]{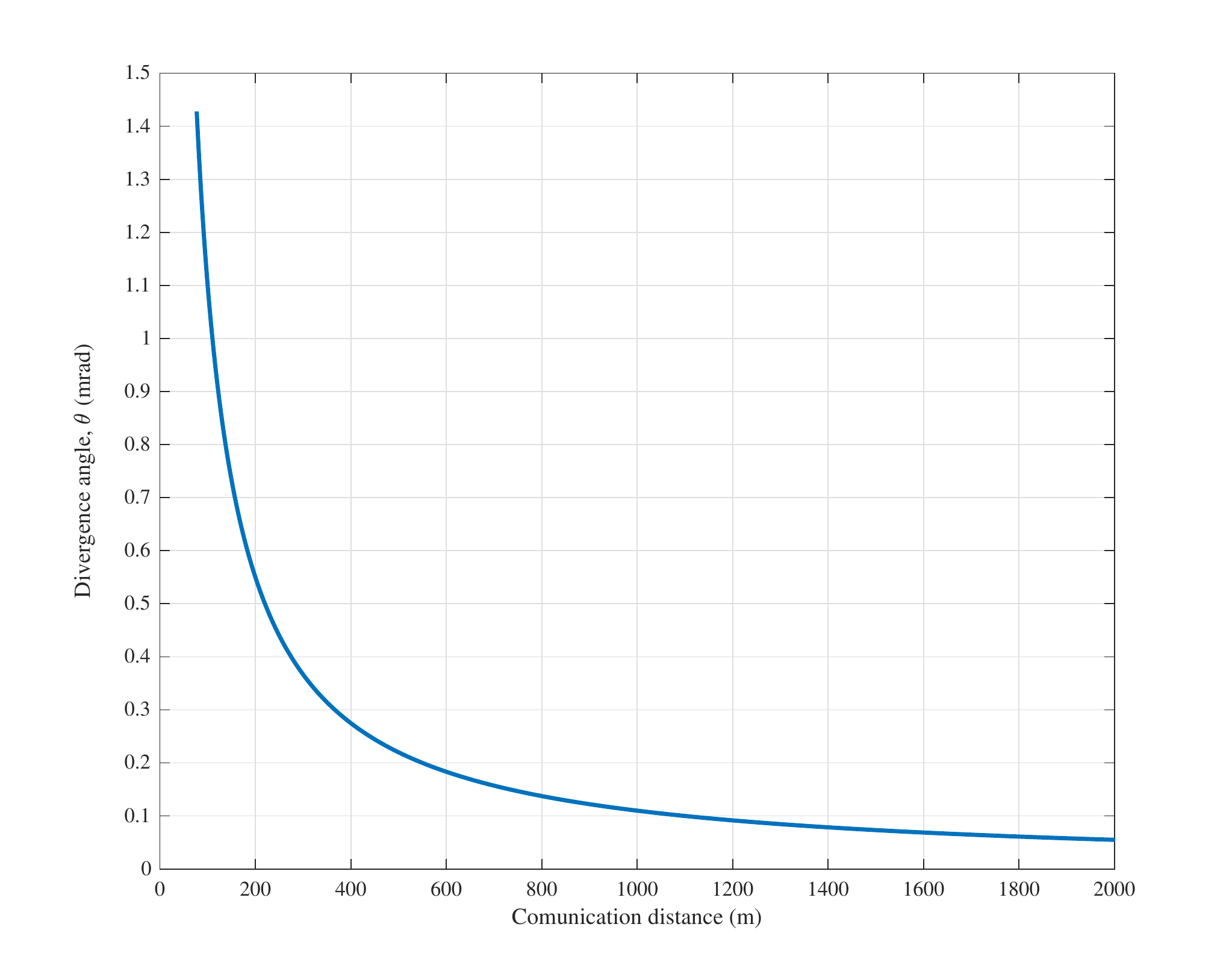}
	\caption{Divergence angle of the proposed adaptive beam as a function of the communication distance between a transceiver on the train and a base station.}
	\label{fig:fullDivAngle_vs_distance}
\end{figure}

Figure \ref{fig:fullDivAngle_vs_distance} shows the divergence angle variation of the proposed adaptive beam as the communication distance between a transceiver on the train and a base station changes. The adaptive beam in this figure adapts its divergence angle to keep the beam width equal to the diameter of the receiver aperture as the communication distance varies. Specifically, having a beam width equal to the diameter of the receiver aperture makes the alignment between the communicating terminals more effective than using a fixed beam having a beam divergence of 1 mrad. Because the fixed beam creates a beam width smaller than the receiver aperture diameter up to a communication distance of 110 m, the adaptive beam is preferred for such a communication distance. Moreover, the width of the fixed beam becomes larger than the receiver aperture diameter as the communication distance goes beyond 110 m. Having a larger beam width than the receiver aperture diameter increases the geometric loss, which results in a decrease of the received power. The adaptive beam, on the other hand, reduces the geometric loss by constantly adapting the beam divergence and the beam width as the communication distance varies. Therefore, an adaptive beam is preferred over a fixed beam in a FSOC system for HSTs as it attains a higher received power for communications distances over 110 m for this specific scenario.

We use 75 m as the shortest communication distance between a train transceiver and a base station in our evaluations because $L_{geo}$ becomes greater than one for a smaller distance, whereas its range should be in [0, 1] for a fixed divergence beam having a divergence angle of 1 mrad. Moreover, we use 2,000 m as the longest communication distance in our evaluations as $\delta$ for the fixed divergence beam becomes negativefor larger distances.

\begin{figure}[t!]
	\centering
	\includegraphics[width=3.5in]{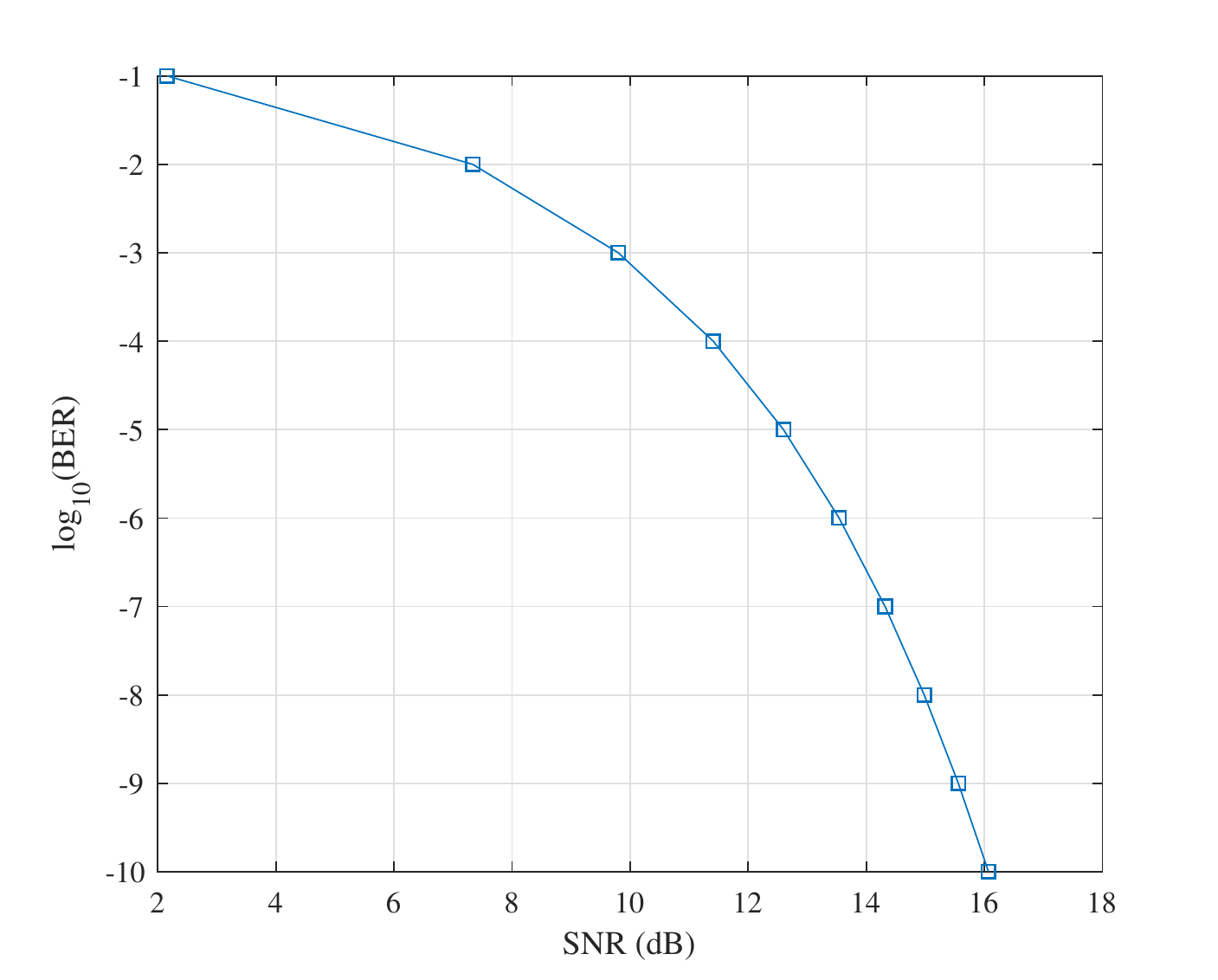}
	\caption{BER as a function of SNR.}
	\label{fig:ber_vs_snr}
\end{figure}

We aim to provide an error-free optical link with a data rate of 1 Gbps between a high-speed train traveling at 400 km/h and the base stations along the track. Therefore, we target a BER equal to or smaller than 10$^{-9}$ in our evaluations. Figure \ref{fig:ber_vs_snr} shows the BER of an intensity-modulation/direct-detection (IM/DD) FSO link as SNR varies according to (\ref{eq:BER}) and (\ref{eq:Qfunc}). Figure \ref{fig:ber_vs_snr} indicates that there is a non-linear relationship between the SNR and the BER. Moreover, Figure \ref{fig:ber_vs_snr} reveals that a BER of 10$^{-9}$ can be guaranteed if the SNR is greater than or equal to 15.56 dB. Therefore, we adopt such SNR as the reference value to calculate the required received power that satisfies our BER requirement according to (\ref{eq:Prx_fog}) and (\ref{eq:apdNoise}). 

\begin{figure}[t!]
	\centering
	\includegraphics[width=3.5in]{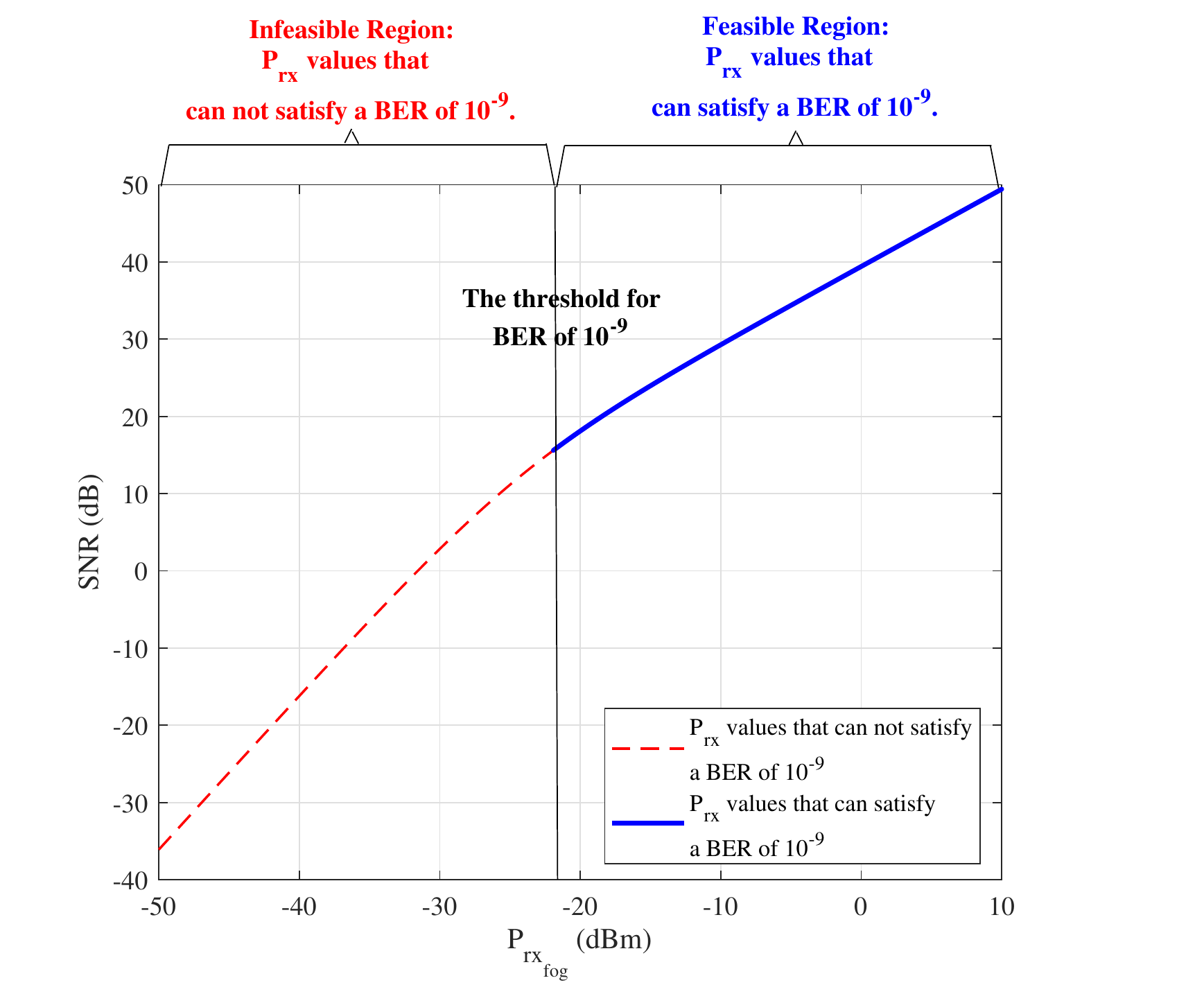}
	\caption{SNR as a function of the received power when a direct detection receiver with an APD is employed. The variance of the noise is calculated according to (\ref{eq:apdNoise}), (\ref{eq:shotNoise}), and (\ref{eq:thermalNoise}). The visibility is 500 m, which corresponds to the presence of moderate fog.}
	\label{fig:prx_vs_snr}
\end{figure}

\begin{figure}[t!]
	\centering
	\includegraphics[width=3.5in]{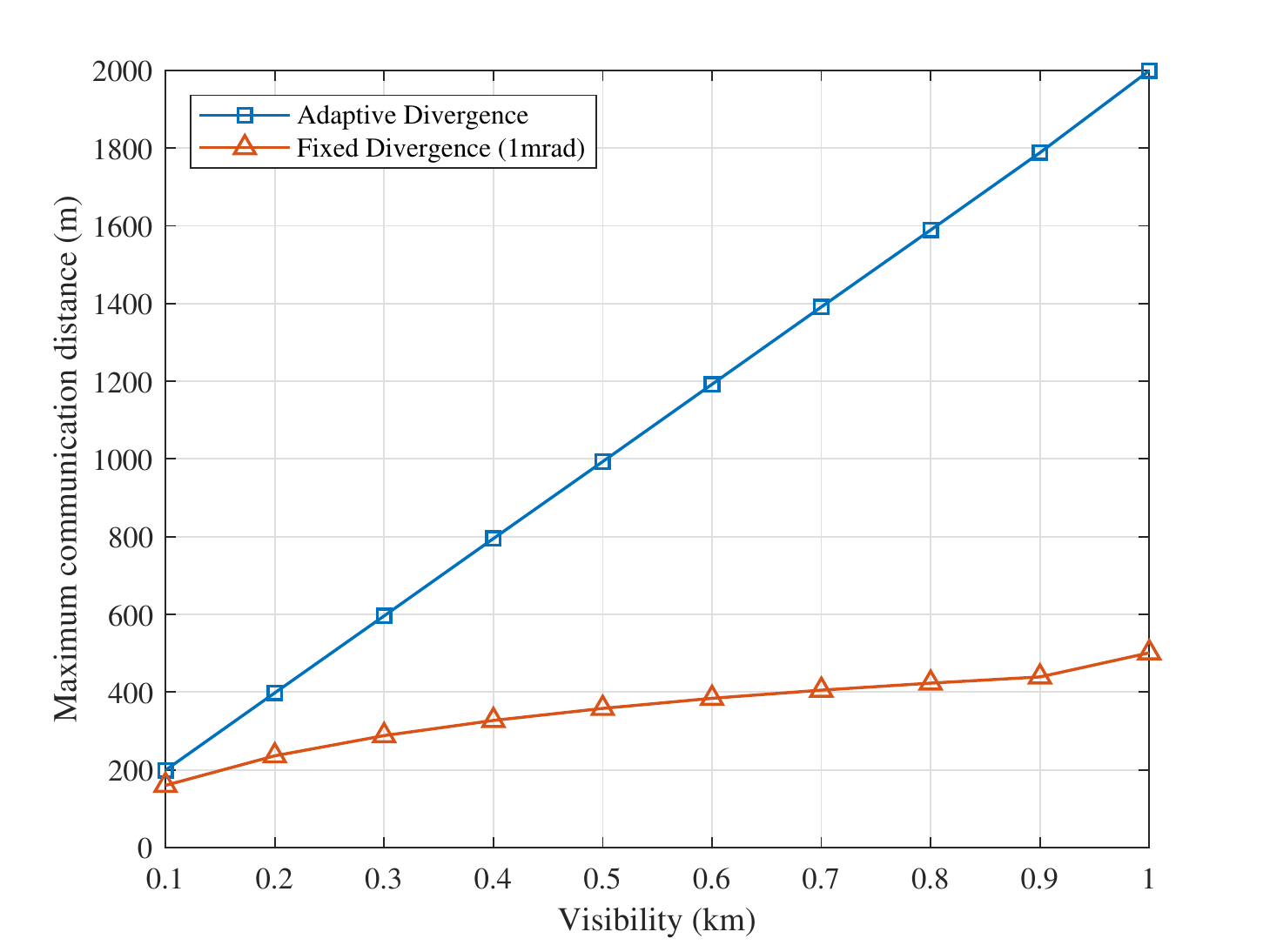}
	\caption{Maximum communication distance for adaptive and fixed divergence angle approaches for different visibilities. A received power of -21.94 dBm is used as the minimum required received power to satisfy a minimum BER of 10$^{-9}$.}
	\label{fig:maxDistance_vs_visibility}
\end{figure}

Figure \ref{fig:prx_vs_snr} shows the SNR of a FSOC system according to (\ref{eq:SNR}) as the received power varies. The variance of the total noise used to calculate (\ref{eq:SNR}) follows (\ref{eq:apdNoise}), (\ref{eq:shotNoise}), and (\ref{eq:thermalNoise}), and the parameters given in Table \ref{table:parameters}. The solid blue line in Figure \ref{fig:prx_vs_snr} indicates the received power values and the corresponding SNR that provide a BER of 10$^{-9}$ or lower. As the received power exceeds -21.94 dBm, the corresponsing SNR becomes greater than 15.56 dB, which yields a BER of maximum 10$^{-9}$, as denoted by the solid blue line in Figure \ref{fig:prx_vs_snr}. Therefore, a received power of -21.94 dBm is used as the minimum required received power when the maximum communication distances of adaptive and fixed divergence beams are calculated and presented in Figure \ref{fig:maxDistance_vs_visibility}. This figure compares the maximum communication distances of adaptive and fixed divergence beams that satisfy a minimum BER of 10$^{-9}$ as the visibility varies. The use of adaptive divergence angle extends the communication distance of a FSOC system three times on average as compared to a fixed divergence angle. 

\begin{figure}[t!]
	\centering
	\includegraphics[width=3.5in]{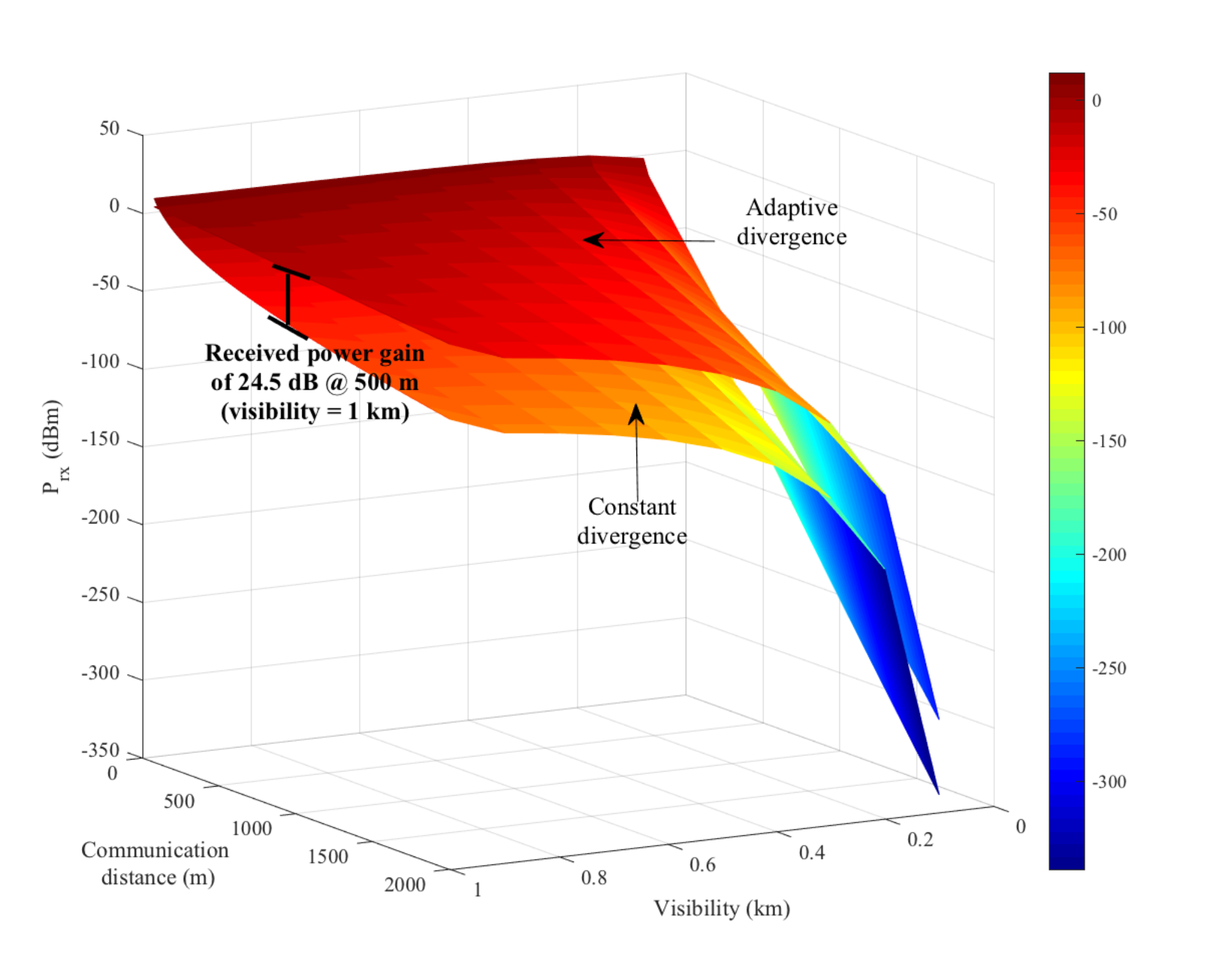}
	\caption{Comparison of the received power for fixed and adaptive divergence beams as a function of the communication distance and visibility.}
	\label{fig:distance_visibility_prx}
\end{figure}

Figure \ref{fig:distance_visibility_prx} shows the impact of meteorological visibility and the communication distance on the received power, in dBm. In this figure, the power loss as a function of visibility is calculated by following (\ref{eq:fog}) and (\ref{eq:fog_q}). The adaptive divergence beam yields higher received power than the fixed divergence beam. Moreover, the received power gap between the adaptive and fixed divergence beams increases as the communication distance increases. For instance, the adaptive divergence beam in this figure yields 33 dB higher received power in average than the fixed divergence beam for a visibility of 1 km.

\begin{figure}[t!]
	\centering
	\includegraphics[width=3.5in]{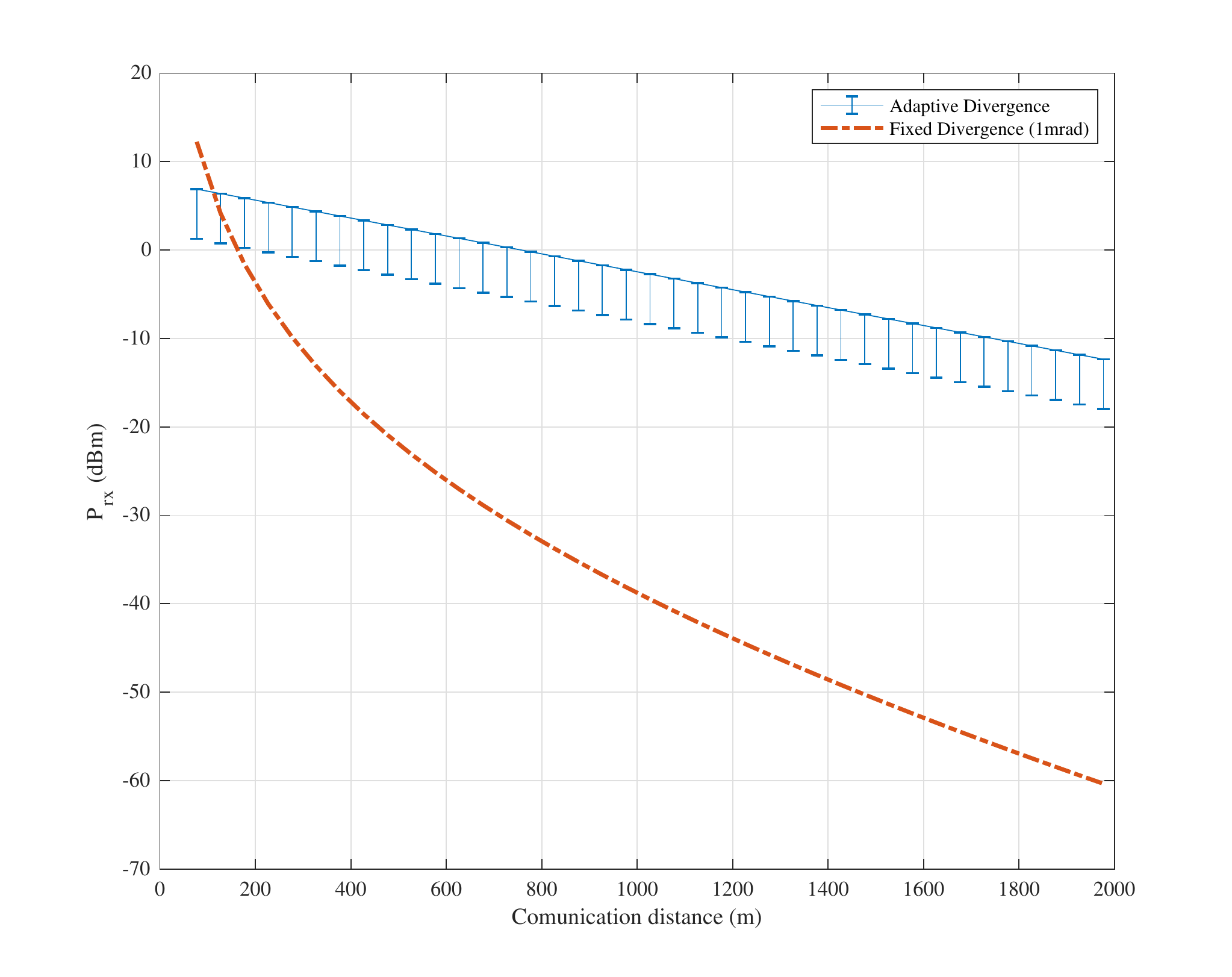}
	\caption{Comparison of the received power between the adaptive and fixed divergence angle approaches. A motorized beam expander with an expansion change time of 5 seconds is used to adjust the beam divergence of the adaptive beam. The divergence angle of the fixed beam angle is 1 mrad. The visibility is 1 km.}
	\label{fig:Prx_Distance_withErrorBars}
\end{figure}

Figure \ref{fig:Prx_Distance_withErrorBars} shows a comparison of received power of the adaptive and fixed divergence beams for a visibility of 1 km. It is assumed that divergence adjustment for the adaptive beam is performed by a motorized beam expander that has an expansion change time of 5 seconds. Note that the expansion change time induces a delay on the divergence adjustment that may create a beam divergence difference between the expected and the actual divergence angles of the transmitted beam at time $t$. The location of the train is periodically sent to the source base station to have the divergence angle of the transmitted beam adjusted by the base station. The exact location of a HST can be detected by track circuits, such as Eurobalises that use the magnetic transponding technology \cite{dhahbi2011study}. The location information can then be disseminated to all base stations by using various communications technologies, such as  global system for mobile communications in railway (GSM-R), universal mobile telecommunications system (UMTS), or satellite \cite{goya2015advanced}. Assuming that the train's location is sent to a source base station at time $t$, the time that the control message carrying the train's location reaches the source base station is $t +  t_{trans} + t_{prop}$, where $ t_{trans}$ and $t_{prop}$ are the transmission and propagation delays for the control message, respectively. As the control message is received by the source base station at time $t +  t_{trans} + t_{prop}$, the communication distance between the train and the base station is calculated based on the location information in the control message. The divergence angle of the transmitted beam is then adjusted according to the calculated communication distance between the train and the base station, and the diameter of the receiver aperture. The time it takes to adjust the divergence angle of the adaptive beam at the base station is $t_{adjust}$. Therefore, the new beam divergence angle for the transmitting beam becomes available at time $t +  t_{trans} + t_{prop} + t_{adjust}$. Because of the small size of the control message (i.e., tens of bytes) and the short communication distances (i.e., hundreds of meters) between the train and base station, $t_{trans} + t_{prop}$ may be considered negligible. The major contributor to the total beam adjustment delay is $t_{adjust}$, which is equal to 5 seconds in this calculation \cite{altechna_motorized_beam_expander, newport_beam_expanders}. Therefore, the beam adjustment is completed at $t + 5$ after the train's location is sent by an Eurobalise at time $t$. The received-power error bars for the adaptive beam in Figure \ref{fig:Prx_Distance_withErrorBars} show the error induced by the total beam adjustment delay by a motorized beam expander. Note that the beam adjustment delay may be eliminated by using an 1xN optical switch, in which an output port is connected to a fixed-magnification beam expander.

\begin{figure}[t!]
	\centering
	\includegraphics[width=3.5in]{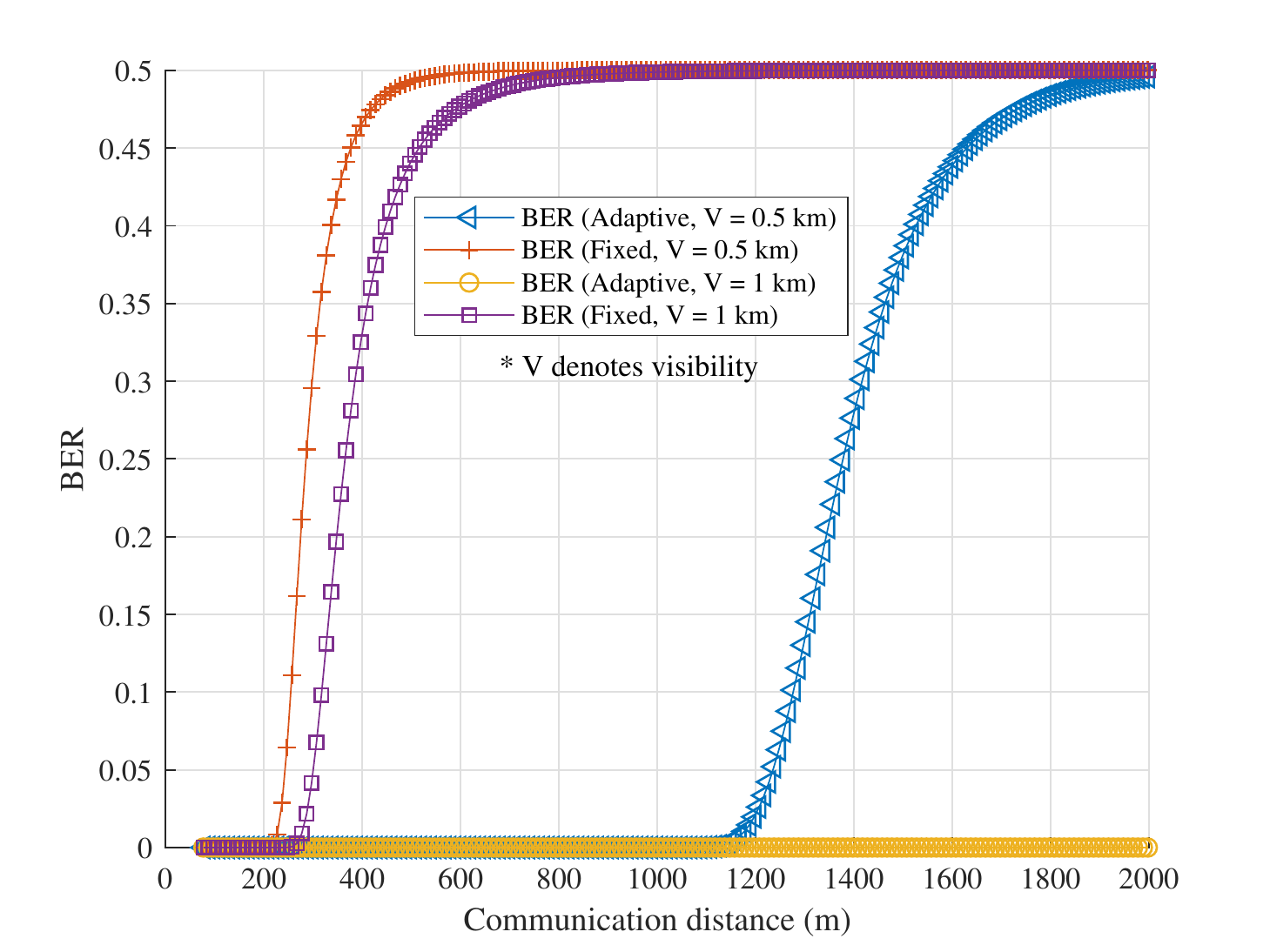}
	\caption{BER of adaptive and fixed divergence beams as a function of the communication distance for visibility values of 0.5 and 1 km.}
	\label{fig:BER_vs_Distance}
\end{figure}

Figure \ref{fig:BER_vs_Distance} shows the BER of the adaptive and fixed divergence beams as a function of the communication distance for visibility values of 0.5 and 1 km. The fixed divergence beam can guarantee a BER of 10$^{-9}$ up to 190 and 224 m for visibility values of 0.5 and 1 km, respectively. The BER of a FSOC system that uses a fixed-divergence beam quickly increases and converges to 0.5 as the communication distances get longer than 190 and 224 m for visibility values of 0.5 and 1 km , respectively. The adaptive divergence beam, on the other hand, extends the communication distances up to 994 and 2,000 m while guaranteing a BER of 10$^{-9}$ for the visibilities of 0.5 and 1 km, respectively. Note that the adaptive beam yields a BER smaller than or equal to 10$^{-9}$ for distances longer than 2,000 m for a visibility of 1 km. Because of the limitations on $\delta$ for the fixed divergence beam, the results in Figure \ref{fig:BER_vs_Distance} do not show communication distances longer than 2,000 m.

\section{Related Works}
\label{sec:relatedwork}
A method that minimizes the transmission power of a building-to-building FSOC system for a given bit error probability (BEP) is presented \cite{arnon2003optimization}. BEP is derived by taking a radial pointing error that represents building-sway statistics into account. The elevation or azimuth standard deviation of radial pointing-error angle is associated to the amplitude of building sway. The divergence angle that minimizes the transmitter power is evaluated for different BEPs when the building sways. The results in \cite{arnon2003optimization} show that using an optimum divergence angle yields a reduction in the transmitter power by more than 4 dB as compared to a system with half or twice of the optimum divergence angle for a BEP of 10$^{-9}$. This work differs from our work as we propose to use an adaptive beam for mobile FSOC, such as HSTs. Moreover, the work in \cite{arnon2003optimization} does not take weather conditions into considiration.

The use of adaptive beam divergence and adaptive transmission power was proposed to increase the likelihood of maintaining an FSO link in case of a transmitter-receiver misalignment for a mobile FSOC system \cite{1563295, lopresti2006adaptive}. The divergence angle of the transmitted beam is controlled by changing the distance between an optical fiber serving as the light source and a collimating lens placed in front of the fiber. The extent of allowable misalignment between the transmitter and receiver is calculated as a function of the transmitter power, divergence angle of the transmitted beam, communication distance, and field-of-view (FOV) of the receiver through simulations and preliminary experiments. The results in \cite{1563295, lopresti2006adaptive} show that the optimum divergence angle used to attain the maximum allowable misalignment decreases as the distance between a transmitter and receiver increases. The results in these works also show that the FOV of receivers imposes an upper limit on the effective transmitter power to achieve the maximum allowable misalignment between a transmitter and receiver. These studies shed light on adaptive selection of the divergence angle and transmitter power to maximize the allowable misalignment between a transmitter-receiver pair. However, calculations presented in \cite{1563295, lopresti2006adaptive} are not directly applicable to HST communication because parameters specific to HST communications, such as the geometry of the train and tracks, and the speed of the HST are not considered.

An adaptive beam divergence technique for inter-UAV FSOC was proposed to minimize geometric and pointing losses at a receiving UAV as the distance between a pair of communicating UAVs varies \cite{heng2008adaptive}. A fixed beam that employs a fixed divergence angle degrades the performance of an UAV-to-UAV FSO link as the communication distance changes. Moreover, UAVs are unable to pinpoint their exact instantaneous positions because of the inherent inaccuracy of their on-board positioning systems. This positioning limitation leads to a spherical error region called uncertainty area. The actual position of an UAV can be anywhere within the uncertainty area. Therefore, each UAV uses an adaptive beam to cover the uncertainty area of its corresponding receiver as the communication distance varies. This  consideration allows the receiving UAV to faster acquire the optical signal emitted by the transmitter. The use of an adaptive divergence angle limits the receiving power loss as the geometric and pointing losses are minimized.

\section{Conclusions}
\label{sec:conclusions}
We have proposed an adaptive beam that adapts its divergence angle according to the receiver aperture diameter and the communication distance to improve the received power and ease the alignment between the communicating terminals as compared to a fixed-divergence beam in a FSOC system for HSTs. Our results showed that the proposed adaptive beam outperforms a fixed-divergence beam that uses a divergence angle of 1 mrad by an average received-power difference of approximately 33 dB. Moreover, the adaptive beam approach increases the maximum communication distance of a FSOC system for HSTs with an average of 742 m over a fixed-beam approach by guaranteeing a BER of 10$^{-9}$ for different visibility values ranging from 0.1 to 1 km. We have also proposed a new placement of ground transceivers on top of the track (above the train passage) of a FSOC system for HSTs; for an optimum alignment with the train movement. The proposed transceiver placement decreases the lateral distance between the transceiver on the train and a base station, which increases the received power of 3.8 dB in average over the base station layout that places the base stations next to track. 


\end{document}